\documentclass{optica-article}

\journal{opticajournal}

\articletype{Research Article}

\usepackage{soul}
\usepackage{lineno}
\usepackage{braket}
\usepackage{graphicx}
\graphicspath{ {./figures/} }
\usepackage{xfrac}

\definecolor{cola}{rgb}{0.7,0.1,0.1}
\definecolor{colb}{rgb}{0.9,0.4,0}
\definecolor{colc}{rgb}{0.3,0.7,0}
\definecolor{cold}{rgb}{0,0.35,0.75}
\definecolor{cole}{rgb}{0.63, 0.13, 0.94}
\definecolor{colf}{rgb}{0.5, 0.5, 0.5}

\newcommand{\Tone}{$T_1$}

\newcommand{\langl}{\begin{picture}(4.5,7)
\put(1.1,2.5){\rotatebox{60}{\line(1,0){5.5}}}
\put(1.1,2.5){\rotatebox{300}{\line(1,0){5.5}}}
\end{picture}}
\newcommand{\rangl}{\begin{picture}(4.5,7)
\put(.9,2.5){\rotatebox{120}{\line(1,0){5.5}}}
\put(.9,2.5){\rotatebox{240}{\line(1,0){5.5}}}
\end{picture}}

\newcommand{\Do}{D\textsuperscript{0}}
\newcommand{\DoX}{D\textsuperscript{0}X}
\newcommand{\DoXs}{D\textsuperscript{0}X\textsuperscript{*}}
\newcommand{\transX}{\Do\,$\leftrightarrow$\,\DoX}
\newcommand{\transXTES}{\Do\,(2s or 2p)\,$\leftrightarrow$\,\DoX}
\newcommand{\transXs}{\Do\,$\leftrightarrow$\,\DoXs}

\newcommand{\up}{$\ket{\uparrow}$}
\newcommand{\down}{$\ket{\downarrow}$}
\newcommand{\uph}{$\ket{\Uparrow}$}
\newcommand{\downh}{$\ket{\Downarrow}$}

\begin{document}

\title{Contributions to the optical linewidth of shallow donor --  bound excitonic transition in ZnO}

\author{Vasileios Niaouris,\authormark{1,\textdagger,*} Samuel H. D'Ambrosia,\authormark{1,\textdagger} Christian Zimmermann,\authormark{1} Xingyi Wang,\authormark{2} Ethan R. Hansen,\authormark{1} Michael Titze,\authormark{3} Edward S. Bielejec,\authormark{3} Kai-Mei C. Fu,\authormark{1,2,4}}

\address{
\authormark{1} Department of Physics, University of Washington, Seattle, Washington 98195, USA \\
\authormark{2} Department of Electrical Engineering, University of Washington, Seattle, Washington 98195, USA \\
\authormark{3} Sandia National Laboratories, Albuquerque,
NM 87123, USA \\
\authormark{4} Physical Sciences Division, Pacific Northwest National Laboratory, Richland, WA 99352, USA \\
\authormark{\textdagger} The authors contributed equally to this work. 
}

\email{\authormark{*}niaouris@uw.edu}

\begin{abstract}
Neutral shallow donors in zinc oxide (ZnO) are spin qubits with optical access via the donor-bound exciton. 
This spin-photon interface enables applications in quantum networking, memories and transduction. 
Essential optical parameters which impact the spin-photon interface include radiative lifetime, optical inhomogeneous and homogeneous linewidth and optical depth. We study the donor-bound exciton optical linewidth properties of Al, Ga, and In donors in single-crystal ZnO. 
The ensemble photoluminescence linewidth ranges from 4-11\,GHz, less than two orders of magnitude larger than the expected lifetime-limited linewidth. 
The ensemble linewidth remains narrow in absorption through samples with an estimated optical depth up to several hundred. 
The primary thermal relaxation mechanism is identified and found to have a negligible contribution to the total linewidth at 2\,K. We find that inhomogeneous broadening due to the disordered isotopic environment in natural ZnO is significant, contributing 2\,GHz. 
Two-laser spectral hole burning measurements, indicate the dominant mechanism, however, is homogeneous. 
Despite this broadening, the high homogeneity, large optical depth and potential for isotope purification indicate that the optical properties of the ZnO donor-bound exciton are promising for a wide range of quantum technologies and motivate a need to improve the isotope and chemical purity of ZnO for quantum technologies.
\end{abstract}

\section{Introduction}

Impurities in solid-state crystals with optical access have received significant attention as potential spin-qubit candidates for quantum memories\cite{heshami2016ome,lvovsky2009oqm}, transduction\cite{han2021moq, lauk2020pqt}, and networks~\cite{awschalom2018qto,wolfowicz2021qgs}, with applications in quantum computing~\cite{benjamin2009qce, weber2010qcd, ladd2010qc}
and quantum communication~\cite{wehner2018qiv, orieux2016rai}. 
Advantageous properties include their potential for scalable device integration~\cite{pelucchi2022pgo}, strong radiative oscillator strength~\cite{lvovsky2009oqm, awschalom2018qto}, spin-dependent optical transitions~\cite{bassett2019qdd}, and long spin coherence times~\cite{bradley2019tqs, zhong2015oan}. In this work we characterize the optical linewidth of the donor-bound exciton transition for ensembles of three shallow donors in ZnO: Al, Ga, and In. 

The neutral shallow donor (\Do) in ZnO, a direct band gap semiconductor, is a spin-1/2 electron qubit system~\cite{mccluskey2009dzo} with the potential for further coupling to the donor nuclear spin. In the presence of a magnetic field, the \Do-bound electron states demonstrate spin-relaxation times up 0.5\,s~\cite{niaouris2022esr} and ensemble coherence times of 50\,\textmu s in natural ZnO~\cite{linpeng2018cps}.
The potential for longer coherence times through isotope and chemical purification~\cite{linpeng2018cps, tribollet2009ten} make \Do\ in ZnO an attractive spin-qubit candidate for photon-based quantum technologies. 

The donor is optically coupled to the donor-bound exciton (\DoX). Two figures of merit for optically-active quantum defects are the oscillator strength, and the ratio of the optical transition linewidth to the Fourier-transform limited linewidth. High oscillator strengths are desirable since they are proportional to the photon emission rate of single photon sources and high optical depth is beneficial for optical quantum memories. In ZnO,  the \DoX\ is bright with a radiative lifetime of 0.86, 1.06 and 1.35\;ns for Al, Ga, and In \DoX\ respectively~\cite{wagner2011bez}. 
A linewidth broader than the lifetime transform limit impacts photon indistinguishability and the strength of the photon-spin interaction, for both ensemble~\cite{lvovsky2009oqm} and single defect applications~\cite{janitz2020cqe}.  

Our goal is to elucidate the various sources - both homogeneous and inhomogeneous - that contribute to the \transX\ transition linewidth beyond this radiative limit. After describing the experimental platform in Sec.~\ref{sec:setup}, in Sec.~\ref{sec:optical_linewidth} we show that the inhomogeneous ensemble optical linewidth of \transX\ at 1.8\,K can be as low as 7\,GHz, compared to the $\mathcal{O}(100\text{\;MHz})$ Fourier-transformed lifetime linewidth.
Transmission measurements also show a very high estimated optical depth of 25 to 300 for the Ga and Al ensembles, %orders of magnitude greater than those in rare earth ion-doped (REI) crystals~\cite{thiel2011red} and silicon radiation damage centers~\cite{higginbottom2023mtp}, and 
approaching that of cold atoms~\cite{geng2014eit,hsiao2014cam}. By measuring the linewidth as a function of temperature in Sec.~\ref{sec:temp_dep}, we find that the dominant phonon contribution to the linewidth is via population relaxation between the \DoX\ excited states. This homogeneous broadening mechanism is negligible at 2\,K.  
In Sec.~\ref{sec:isotope}, we calculate an intrinsic inhomogeneous broadening due to isotopic variation on the order of a few GHz. Finally, in Sec.~\ref{sec:rSHB}, we probe the homogeneous linewidth via spectral anti-hole burning. We find surprisingly that the spectral anti-hole linewidth, which is measured on microseconds timescales, is the dominant contributor to the ensemble linewidth. We conclude with a discussion on how these properties may impact donor qubit operation and how they may be further improved.

\section{Experimental Setup}
\label{sec:setup}
The two samples studied in this work are 300\,\textmu m-thick Tokyo Denpa single-crystal substrates from the same parent crystal. 
Sample A is untreated and Sample B has undergone indium implantation and annealing to form In donors at the surface ($\sim$200\,nm deep)~\cite{wang2023pdq}. 
Three donor species are studied: Al, Ga, and In substituting for Zn. The donor concentrations in the first two microns of sample B's surface were determined by secondary ion mass spectroscopy (SIMS) measurements as 1.2$\cdot10^{15}$\,cm$^{-3}$ for Al and 9.2$\cdot10^{15}$\,cm$^{-3}$ for Ga~\cite{wang2023pdq}. 
The bulk doping concentration for In was below the SIMS detection limit. 
We note that the PL intensity of the individual lines can vary by an order of magnitude across the sample. One possible cause would be non-uniform incorporation during growth. 

The samples are mounted in a helium immersion cryostat with a superconducting magnet. 
The magnetic field $\vec{B}$ direction is fixed.
The [0001] sample surface is perpendicular to crystal axis $\hat{c}$.
We access different magnetic field geometries by rotating the sample to either the $\vec{B}\parallel\hat{c}$ window (Faraday geometry) or $\vec{B}\perp\hat{c}$ window (Voigt geometry).
The optical axis $\hat{k}$ is parallel to the crystal axis $\hat{c}$ in all measurements. 
The optical path for each experiment in this manuscript and equipment part numbers used for these experiments are detailed in the supplemental material Sec.~S1.

\begin{figure*}[t]
    \centering
    \includegraphics[width=0.97\textwidth]{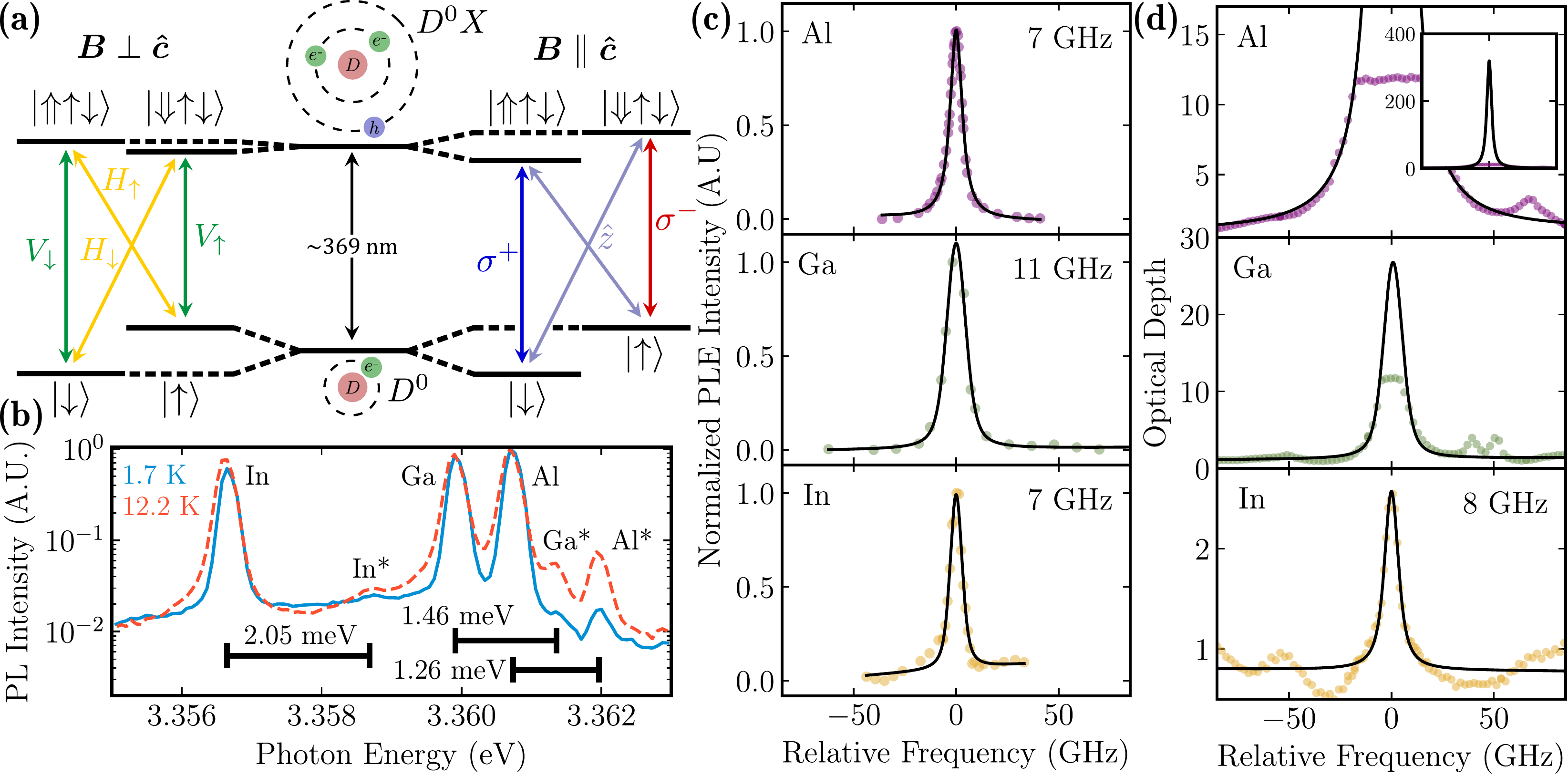}
    \caption{\label{fig:intro} 
    (a) Energy diagram and selection rules in Voigt (left) and Faraday (right) geometries. 
    $V_x$ and $H_x$ denote vertically and horizontally polarized light with $x$ denoting the \Do\ ground electron-spin state.
    Circular polarizations are denoted as $\sigma^+$ and $\sigma^+$, while $\hat{z}$ corresponds to linear polarization parallel to the optical axis. 
    (b) PL spectrum of Sample B. Al, Ga, and In (Al$^*$, Ga$^*$ and In$^*$) label the corresponding \transX\ (\transXs) transitions. Excitation energy is 3.44\,eV, power is 30\,nW and diameter is 600\,nm. 
    (c) PLE of the three donor species at 1.7\,K for Sample A. Fits are to a Voigt profile and take into account incident power oscillations due to a beam-splitter (Sec.~S2). Excitation power is 200\,nW for Al, 100\,nW for Ga and 1.15\,\textmu W for In. The beam diameter is 600\,nm.
    (d) Optical density for the three donor species at 1.7\,K in Sample A. Fits are to a Voigt profile. For Al and Ga, we constrain the FWHM to the FWHM found in the PLE measurements in (c). The calculation of OD from transmission is given in Sec.~S3. Excitation power is 15\,nW with a diameter of 380\,\textmu m. PL spectra in this experimental configuration are given in Sec.~S4.}
\end{figure*}

The spin-1/2 neutral shallow donor electron (\Do) is optically coupled to the donor-bound exciton (\DoX) state, where an electron-hole pair is bound to the neutral donor. In an applied magnetic field, the spin-1/2 electron of \Do\ splits to \down\ and \up. The electron g-factor of the donor is nearly isotropic with $g_e=1.97$~\cite{linpeng2018cps}. The \DoX\ also splits in applied field due to the bound hole spin \uph\ and \downh. The hole g-factor is highly anisotropic~\cite{meyer2004bed} ranging from -1.2 in Faraday to 0.3 in Voigt~\cite{linpeng2018cps, niaouris2022esr}.
We can optically access and manipulate the two \Do\ spins via two $\Lambda$-systems (Fig.~\ref{fig:intro}(a)). In the Voigt geometry, the relative strength of the transitions is nearly identical. In the Faraday geometry, the branching ratio is 99:1 between the $\sigma^\pm$ and $z$-polarized transitions~\cite{linpeng2020dqd, lambrecht2002vbo}.

\section{Results}
\label{sec:results}

\subsection{Ensemble Optical Linewidth}
\label{sec:optical_linewidth}

A photoluminescence (PL) spectrum showing emission from all three donors in sample B is shown in Fig.~\ref{fig:intro}b. We observe six peaks between 3.355\,eV and 3.363\,eV. The transitions labelled Al, Ga, and In correspond to the transition between the 1s \Do\ and the lowest energy \DoX\ state of the respective donor species (\transX)~\cite{wagner2011bez}. 
The transitions Al$^*$, Ga$^*$, and In$^*$ correspond to transitions from an excited \DoX\ state, which we denote as \DoXs. The \DoXs\ intensity
becomes brighter with increasing temperature. The energy differences between the \DoXs\ and \DoX\ for Al, Ga, and In are 1.26\,meV, 1.46\,meV, and 2.05\,meV respectively. As discussed in Sec.~\ref{sec:temp_dep}, the presence of these excited states contribute to a phonon-induced broadening of the \DoX\ transitions at elevated temperatures.

The linewidths in Fig.~\ref{fig:intro}b are spectrometer resolution-limited (55\,GHz). 
We thus utilize micro-photoluminescence excitation (PLE) spectroscopy to determine the ensemble linewidths. 
In PLE measurements, we scan a continuous-wave laser near the \transX\ resonance and sum over the collected side-band PL. 
The excitation laser is measured by a wavemeter with a resolution of 0.5\,GHz.
The side-band PL consists of the two electron satellite (TES) \transXTES, first and second phonon replicas (1LO and 2LO) and first phonon replica of the TES (1LO-TES), as described in prior work~\cite{niaouris2022esr, wang2023pdq}. 
The measured PLE linewidths of the Al, Ga, and In \DoX\ transitions are $7.1\pm0.1$\,GHz, $11.1\pm0.3$\,GHz, and $7\pm0.3$\,GHz, respectively. 

The lifetime-limited linewidth $\Gamma$ is given by $\Gamma=1/(2\pi \tau)$, where $\tau$ is the lifetime measured in the literature~\cite{wagner2011bez}.
The Al and Ga lifetimes contain a non-radiative component which is attributed to a non-radiative surface recombination mechanism~\cite{chen102ddb} but could also be due to exciton dissociation. 
Using the faster component as the lifetime value, the lifetime-limited linewidths are 0.5\,GHz for Al, 0.4\,GHz for Ga, and 0.1\,GHz for In.
The PLE linewidth is almost two orders of magnitude larger than the expected lifetime limit. 
The observed broadening could be due to homogeneous factors that would affect a single center on the timescale of our measurements including phonon-broadening, spectral diffusion and hyperfine interactions. 
It could also be due to inhomogeneous factors such as static microscopic electric and strain fields and isotope disorder.
While significantly broader than the lifetime limit, the PLE ensemble linewidth is still remarkably narrow and is less than 100 times the lifetime limit. 
In comparison, the best ratio of inhomogeneous:radiative linewidths for {\it in-situ} doped nitrogen vacancy (NV) centers is 1000~\cite{santori2010nqo}; this ratio is even larger for REIs\cite{thiel2011red}. 

This narrow linewidth persists in transmission measurements.  
Fig.~\ref{fig:intro}d shows the optical depth (OD) $\alpha d$ through the $d=300$\,\textmu m substrate, where 
$\alpha$ is the frequency-dependent absorption constant. 
The In \DoX\ absorption linewidth through the 300\,\textmu m-thick sample is only 10\,\% broader than the micro-PLE linewidth, suggesting that high optical homogeneity persists over large volumes. 
The Al and Ga \DoX\ transmission measurements saturate at 11.5 OD. This saturation occurs when the resonant PL intensity from the sample exceeds the transmitted laser power. Assuming that the linewidth does not significantly vary between micro-PLE and transmission measurements, a fit to the wings of the transmission spectra give a peak OD of 25 for Ga and more than 300 for Al. The area under each OD peak is proportional to the number of donors in the probed ensemble. 
 We estimate the average donor density for each donor species; $N_{\rm Al} = 7.5\cdot 10^{15} \rm\, cm^{-3}$, $N_{\rm Ga} = 9.9\cdot 10^{14} \rm\, cm^{-3}$, and $N_{\rm In} = 7.4\cdot 10^{13} \rm\, cm^{-3}$ (Sec.~S5). These values, measured for Sample A, are within an order of magnitude of the SIMS values measured for sample B. The relative concentrations are consistent with the PL spectra for this sample (Sec.~S4). 

An optical depth $\alpha d \gg 1$  is a critical requirement for the development of efficient optical quantum memories~\cite{lvovsky2009oqm}. Warm atomic vapors~\cite{phillips2008ols} and cold atoms\cite{sparkes2013uho} can achieve optical depths from the tens up to a thousand in cm-scale devices. These high optical depths are more challenging to achieve in solid-state systems which typically have much lower oscillator strengths~\cite{thiel2011red, higginbottom2023mtp} and large inhomogeneous broadening. In comparison, the ZnO donor-bound exciton system combines high oscillator strength, high homogeneity, and optical depth at residual donor densities. 

\subsection{Temperature-dependent phonon broadening}
\label{sec:temp_dep}

\begin{figure}[t]
    \centering
    \includegraphics[width=1\textwidth]{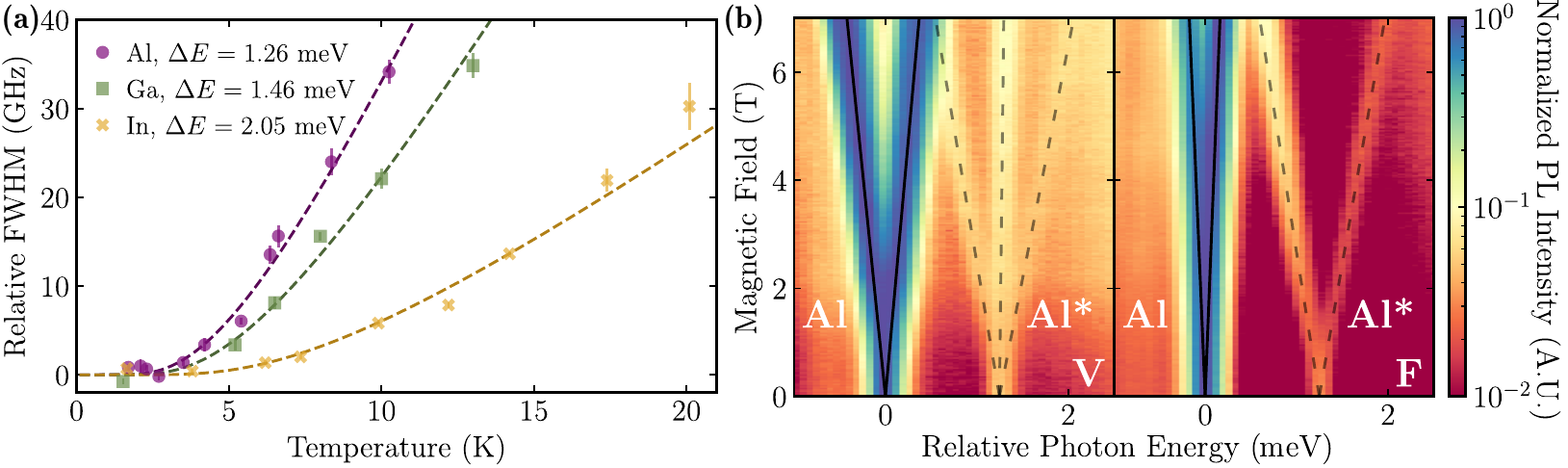}
    \caption{\label{fig:doxstar}
    (a) PLE linewidth as a function of temperature for Al, Ga, and In. Each data-set is shown with the 0\,K linewidth subtracted. 
    All data-sets are fit to Eq.~\ref{eq:fwhm_temp}, with $\Delta E$ determined from Fig.~\ref{fig:intro}b. $\Delta \nu_0 =7.4 \pm 0.4$\,GHz, $11.8 \pm 0.8$\,GHz, and $6.5 \pm 0.5$\,GHz and $a = 110 \pm 5$\,GHz, $99 \pm 6$\,GHz, and $ 59 \pm 4$\,GHz, for Al, Ga, and In respectively. 
    (b) Magneto-PL of Al donor in Sample A. (left) Voigt (T=7.4\,K) (right) Faraday (T = 5.5\,K). Excitation at 3.44\,eV. In Voigt, the splitting of the (unresolved) doublets corresponds to $g_e = 1.95$, while the \DoXs\ splitting corresponds to effective g-factors of 1.87 between observed transitions. In Faraday, the \DoX\ splitting corresponds to  an exciton g-factor $g_{e-h} = 0.83$. The \DoXs\ splitting corresponds to an effective g-factor of 3.39.
}
\end{figure}

Defect-phonon interactions can be a dominant homogeneous optical dephasing mechanism, hence we need to confirm that we are performing our linewidth study below thermal broadening limit. %Defect-phonon interactions are typically considered a homogeneous broadening mechanism due to the fast timescale. 
Fig.~\ref{fig:doxstar}a depicts the PLE linewidth dependence of the \transX\ transition for temperatures varying from 1.5\,to 18\,K for all three {\it in-situ} doped donors in sample A. Implanted In (sample B) at low implantation doses follows a similar dependence, however a stronger temperature dependence is observed at higher doses (Sec.~S6).

The onset and rate of the temperature-dependent linewidth broadening is different for the different donor species. The observed temperature dependence can be modeled by a single phonon absorption process with rate $\Gamma(T)$ from the lowest-energy \DoX\ to an excited \DoXs\ state. In ZnO, the primary interaction is likely the piezo-phonon interaction, as has been observed in longitudinal spin relaxation ~\cite{niaouris2022esr}. The precise phonon absorption rate is unclear due to the lack of a satisfactory model for \DoXs, however, the rate of phonon absorption is proportional to the phonon number $N_{ph}$ at energy $\Delta E$, the energy splitting between the \DoX\ and \DoXs states. $N_{ph}$ will follow a Bose-Einstein distribution. This allows us to express the total linewidth as 
\begin{equation}
    \label{eq:fwhm_temp}
    \Delta \nu (T) = \Delta \nu_0 + a N_{ph}(T) = \Delta \nu_0 + a \left(e^{\Delta E/k_B T} - 1\right)^{-1},
\end{equation}
in which $\Delta \nu_0$ is the temperature-independent component of the linewidth, and $aN_{ph}(T)$ is the broadening due to excitation from \DoX\ to \DoXs, and $a$ is a scaling factor that is donor-dependent. %For each curve in Fig.~\ref{fig:doxstar}a, the minimum measured linewidth $\Delta \nu_0$ is subtracted for ease of comparison. 
While this model is valid for low temperatures, it will eventually break down near 60\,K, where the donor-bound exciton starts dissociating to a neutral donor and a free exciton~\cite{hamby2003tde}.

%As depicted in Fig.\ref{fig:intro}a, in our measurements we observe a single prominent \DoXs\ state for each donor. 
A fit to Eq.~\ref{eq:fwhm_temp} using the experimentally measured $\Delta E$ (Fig.\ref{fig:intro}a) shows good agreement to the model (Fig.~\ref{fig:doxstar}a) at least up to 20\,K. 
The presence of the excited state places a fundamental limit on the maximum temperature at which indistinguishable photons can be obtained from an emitter. 
In the absence of Purcell enhancement, the temperature at which the temperature component of the linewidth becomes equal to the radiative limit, i.e. $\Delta\nu_\text{\,T} = \Delta\nu_{\text{rad}}$ is $T = 2.7, \,3.1, \,3.8$\,K for Al, Ga, and In respectively. 
We note that at the lowest temperatures in this study, T = 1.7\,K, the phonon contribution to the linewidth is negligible, $\Delta\nu_{\left(\text{T}\,=\,1.7\,\text{K}\right)} =  20, \,4.6, \,0.05$\,MHz for Al, Ga, and In respectively. 
A difference in the scaling factor $a$ is observed between donors. 
This difference may be due to the difference in the polarizability of the donors~\cite{niaouris2022esr, khaetskii2001sft}, which inversely depends on state localization~\cite{szabo2022fds}. 
Polarizability will increase from In to Ga to Al, which is consistent with the observed increase in $a$.

In order to gain insight into the nature of the \DoXs\ state, we collected PL spectra as a function of magnetic field. Fig.~\ref{fig:doxstar}b depicts the Al \DoX\ and \DoXs\ field dependence. Sec.~S7 includes magneto-PL for all donors. In both geometries, the \DoX\ observed splittings are consistent with the reported g-factors for the electron and hole
%i.e. an electron g-factor of $g_e=1.97$, and hole g-factors of $g_h^{\perp}=0.34$ and $g_h^{\parallel}=-1.22$ for Voigt and Faraday geometries respectively
~\cite{linpeng2018cps, niaouris2022esr}. However we observe three transitions for \DoXs\  in Voigt and a very large exciton splitting ($g_{\text{exciton}} = 3.39$) in Faraday. Our g-factor measurements differ from Ref.~\cite{meyer2010esp} in which it was reported that the g-factor of \DoXs\ is the same as that of \DoX. We attribute this discrepancy to the high density of overlapping lines around the Ga and Al \DoXs\ energies in Ref.~\cite{meyer2010esp} which could result in state misidentification. The origin of the three transitions (versus 2 or 4 observed for \DoX) and large g-factor is currently unknown. Interpretation of the excited states of the bound exciton is discussed further in Sec.~S9.

\subsection{Inhomogeneous Broadening Due to Isotopic Composition}
\label{sec:isotope}
After thermal homogeneous broadening, inhomogeneous broadening is often another dominant broadening source.
Inhomogeneous broadening of the \transX\ transition energy can be caused by extrinsic factors such as local variations in strain and electric fields due to point and extended defects. Intrinsic contributions to inhomogeneous broadening can occur due to nuclear spin or isotopic mass composition of the emitter's environment. 
The latter is a dominant inhomogeneous broadening mechanism in high quality natural silicon~\cite{karaiskaj2001pip, karaiskaj2003ora, yang2009hlp}. Here we estimate the effect of isotopic mass composition on the ZnO \DoX\ optical linewidth. The effect of the nuclear spin environment on the linewidth was also considered, however, with the exception of a large In donor hyperfine splitting \cite{wang2023pdq}, this was not found to have a significant effect (Sec.~S10). 

The local isotopic mass environment can effect the \transX\ transition by local variation in the zinc and oxygen isotopes in the defect's environment, by or variation of the isotope of the impurity atom. The \transX\ transition closely follows the local band gap, which is determined in part by the zero-point electron-phonon renormalization~\cite{manjon2003zws, cardona2000iep}. 
This zero-point renormalization energy depends on the average mass of atoms present in the local environment~\cite{cardona2000iep}. To assess the effect of isotopic substitution on a specific state or set of states, the band gap variation must be decomposed into conduction band and valence band shifts~\cite{karaiskaj2002iag, stegner2010iee}. These have different temperature dependences, and thus different zero-point renormalization energies~\cite{cardona2000iep}. In ZnO, the valence band will shift by approximately 80$\%$ of the total band gap shift, while the conduction band exhibits a shift in the opposite direction of 20$\%$~\cite{gorkavenko2007ctd}.

To quantify the resulting shifts, we follow the method presented for silicon in Ref.~\cite{karaiskaj2003ora}, modified for ZnO. Here the carrier wavefunction is discretized on atomic sites $\vec r_i$. The model for the effective mass envelope functions of the \Do\ electron and \DoX\ electrons and hole are given in Sec.~S8. In this model, the energy shift for a given carrier state due to perturbation of its isotopic environment is given by $\langl \Phi_{\text{S,c}}(\vec{r}_i) | H_i^{\text{iso}} | \Phi_{\text{S,c}}(\vec{r}_{i'}) \rangl = \delta_{i,i'}W_{i, c}$,
where $\ket{\Phi_\text{S,c}\left(\vec{r_i}\right)}$ refers to the Bloch function for state S (\Do\ or \DoX), carrier c (e = electron or h = hole), the lattice site $i$ at a distance $\vec{r_i}$ from the impurity, $H_i^{iso}$ is the perturbation term, and $W_{i, c}$ is the energy shift due to the isotopic variation. The energy shift results from a shift in the top of the valence band or bottom of the conduction band, depending on if the carrier is a hole or an electron respectively. Shifts are relative to the lowest mass isotope. The values for these shifts are listed in Table~S4 in Sec.~S12.

\begin{figure}[t]
      \centering
      \includegraphics[width=1.\textwidth]{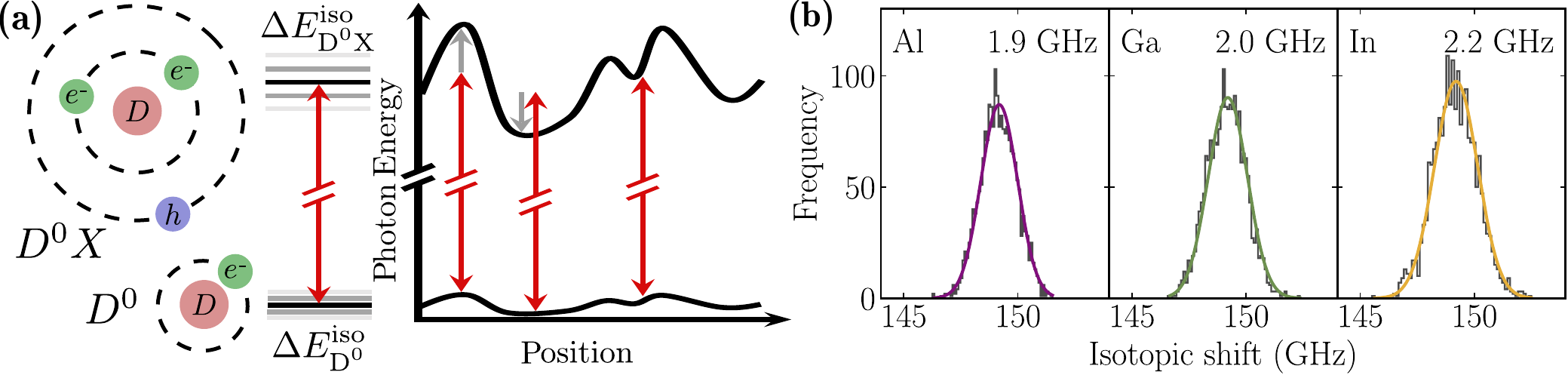}
      \caption{
       Isotopic broadening for each shallow donor type. Different local environments will shift the energies of the \Do\ and \DoX\ by different but correlated amounts, with the \DoX\ shifting more due to the mobility of the valence band in ZnO. We calculate 1.9 GHz broadening for Al, 2.0 GHz for Ga, and 2.2 GHz for In.}
       \label{fig:isotope_effect} 
\end{figure}

The total shift $\Delta E^{iso}_{S,c}$ on each state $S$ and carrier $c$ is
\begin{equation}
    \begin{split}
        \Delta E^{iso}_{S,c} = 
        \Delta E^{imp}_{S,c} +
        & \sum_{i \text{ } \in \text{ lattice sites}} 
        \langl \Psi_{S,c} (\vec{r}_i) |H_{i}^{iso}| \Psi_{S,c}(r_i)\rangl 
    \end{split},
\end{equation}
where $\Psi$ refers to a carrier state with an effective mass envelope function as defined in Sec.~S8. $E^{imp}_{S,c}$ is the shift from substitution of the impurity atom isotope, which was found to have only a small effect, as discussed in Sec.~S11.
The total shift of \transX\ is determined by the difference between the \Do\ and the \DoX\ shifts. 
Since they inhabit the same isotopic environment, their shifts will be correlated (Fig.~\ref{fig:isotope_effect}a). 
The total transition shift is given by $\Delta E_{iso} = (2 \Delta E^{iso}_{D^0X,e} + \Delta E^{iso}_{D^0X,h}) - \Delta E^{iso}_{D^0,e}$, 
accounting for the two electrons and hole in the \DoX\ state, and electron in the \Do\ state.

The isotopic mass environment broadening is determined numerically using 2000 simulated environments for each \Do\ and \DoX. 
As shown in Fig.~\ref{fig:isotope_effect}b, the estimated total contribution of the isotopic mass environment to the inhomogeneous broadening is 1.9\,GHz, 2.0\,GHz, and 2.2\,GHz for Al, Ga, and In respectively. 
Less localized states ({\it e.g.} Al) are affected by a larger number of environmental lattice sites, leading to smaller deviations %between the mean isotopic mass environment and 
in the local isotopic mass environment. 
%which affects the transition energy. 
Using this model for the phosphorus \transX\ transition in natural silicon, which is known to be isotopically broadened, yields an inhomogeneous linewidth that is in good agreement with observation (Sec.~S12) \cite{yang2009hlp}. Thus in the current samples, while not negligible, the isotopic environment is not the dominant broadening mechanism in the ensemble linewidth.   

\subsection{Homogeneous Spectral Anti-Hole Linewidth}
\label{sec:rSHB}

We thus probe the homogeneous linewidth of the Al donor with spectral anti-hole burning measurements to elucidate the nature of the dominant broadening mechanism. The anti-hole linewidth will not be affected by the static varying isotopic environment~\cite{yang2009hlp}.  We perform these two-laser experiments in a pump-probe configuration to probe time-dependent mechanisms such as spectral diffusion. We have also performed continuous-wave (cw) measurements which yield similar linewidths (Sec.~S13). %Here, we study the Al donor which has an expected lifetime-limited linewidth of 0.5\,GHz.

\begin{figure*}[t]
    \centering
    \includegraphics[width=1.0\textwidth]{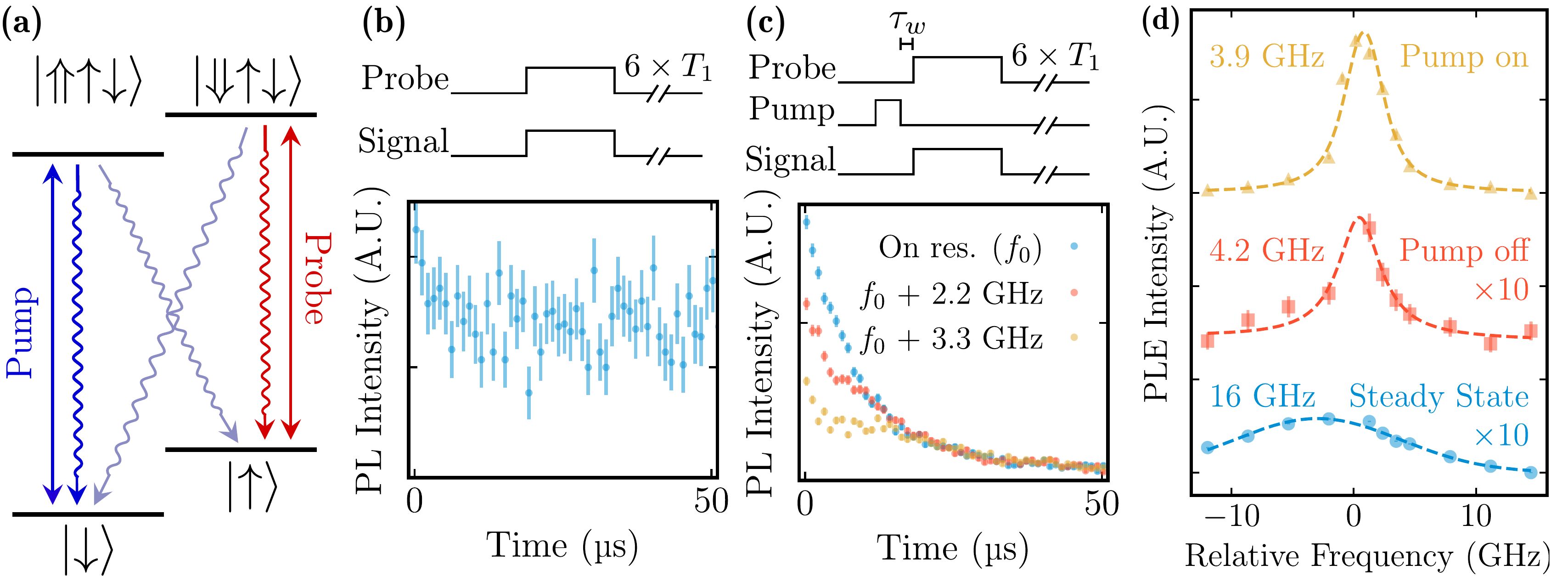}
    \caption{\label{fig:transient} 
    %All depicted data sets are taken in the 
    (a) An energy diagram for spectral hole burning experiment.
    (b) Single laser optical pumping (OP) curve with  230\,nW probe laser power. 
    (c) OP curves in pump-probe experiment with the 230\,nW probe laser, 440\,nW pump power and  100\,\textmu s pump pulse length. Three  excitation frequencies are shown (on resonance,  +2.2\,GHz, and +3.3\,GHz detuned).
    Both (b) and (c) contain a schematic of the pulse sequence. Between cycles, the wait time is $6\times$ the 1.5\,ms longitudinal spin relaxation time (\Tone). 
    (d) PLE curves for transient experiments. Fit is to a Voigt profile. 
    For clarity, each spectrum has been offset vertically, and the steady state and single laser initial state curves are scaled $\times 10$. Faraday geometry at 7\,T and 1.8\,K.
    }
\end{figure*}

Setting a single probe laser resonant to the $\sigma^-$ transition, as shown in Fig.~\ref{fig:transient}a, we optically pump (OP) the sub-ensemble population from the \up\ to the \down\ spin state. Hence, when collecting the side-band emission, we observe a decrease in signal that is proportional to the population depletion of the \down\ spin state (Fig.~\ref{fig:transient}b). In the absence of a pump beam, the probe signal is small due to the small thermal population in the \up\ state.

When we perform optical pumping as a function of probe laser wavelength, intensity at the start of the optical pumping curve varies as the laser is scanned over the resonance. If we integrate this signal in the first 2 microseconds and plot the intensity as a function of probe frequency, we observe a linewidth of $4.2\pm 2.8$\,GHz. This is comparable to the 7\,GHz 0-field cw PLE linewidth. If instead we integrate the signal at the end of the optical pumping curve, {\it i.e.} when the system is in the optically-pumped state, a linewidth of $16.0\pm 0.9$\,GHz is observed (Fig.~\ref{fig:transient}d). The broader linewidth in the optically-pumped state is expected as efficient optical pumping on-resonance decreases the peak intensity. The similarity between the zero-field cw measurements and the time-dependent PLE in which the start of the optical-pumping curve is integrated is expected as optical pumping should not occur at zero-field. 

We next introduce a second excitation laser resonant to the $\sigma^+$ transition (``Pump'' laser in Fig.~\ref{fig:transient}a). The pump laser initializes the sub-ensemble with which it is resonant from the \down\ to the \up\ spin state. After a wait time $\tau_w$, the probe laser optically pumps the sub-ensemble \up\ to the \down\ spin state. In Fig.~\ref{fig:transient}c an enhanced optical pumping signal is observed when the probe laser is applied after the pump. The amplitude of the optical pumping signal, plotted as a function of probe frequency, determines the linewidth of the resonantly pumped sub-ensemble and thus the homogeneous linewidth of the \transX\ transition. As shown in Fig.~\ref{fig:transient}d, we observe a linewidth of $3.9\pm 0.5$\,GHz. Varying the wait time does not affect initial amplitude of the OP curve~(Sec.~S14).
Thus the homogeneous linewidth appears to be the dominant component to the spectral anti-hole linewidth.https://www.overleaf.com/project/6553e95bb022f52101c80712

The observed spectral anti-hole linewidth can be attributed to the \DoX\ state; two laser coherent population trapping experiments, which probe the \Do\ coherence report a two-photon spin linewidth in the tens of MHz for Al~\cite{wang2023pdq}. The time dynamics of the optical pumping curve (Fig.~\ref{fig:transient}c) suggest a laser-induced diffusion process is occurring. In contrast to single laser experiments, in which the optical pumping curve can be characterized by a single exponential~\cite{linpeng2018cps}, complex temporal dynamics of the curve are observed following the pump pulse. As both the intensity and frequency of the probe pulse are fixed, this temporal dynamics suggest a time-dependent change in the transition frequency/lineshape of the probed donors. We expect spectral diffusion is also occurring during the pump pulse, {\it i.e.} as \DoX\ population is depleted at the resonant pump frequency, near-in-frequency \DoX\ population diffuse into this depletion region.  Future studies are required to elucidate the origin of the spectral diffusion process. Potential sources include changes in the microscopic charge environment due to nearby impurities and defects, and instantaneous spectral diffusion (ISD)~\cite{thiel2014mae}, in which excited \DoX\ within the ensemble interact electromagnetically. This is a well-documented phenomenon in REIs where the interaction is dipolar. In the \Do\ case, the effect may be more severe. The \Do\ and \DoX\ effective mass states are significantly extended as compared to REI's with a significant difference between the single electron \Do\ and three-carrier \DoX\ wavefunctions (Sec.~S8). 

\section{Discussion and Outlook}
\label{sec:outlook}

Further studies are required to confirm the origin of the homogeneous linewidth. If it is due to ISD, there are two immediate impacts on quantum information technologies. First, for single quantum defect applications, lower donor densities are required. In REI's, this can be achieved by burning a large spectral hole from which a narrow, low-density anti-hole can be established~\cite{kinos2022mti}. This strategy is only feasible when the homogeneous linewidth is much narrower than the inhomogeneous linewidth, a condition not satisfied in our samples. Thus lower donor dopant densities will be required. On the flip side, a large ISD linewidth indicates a large \DoX-\DoX\ interaction which long-term could provide a mechanism for \Do-\Do\ gates~\cite{longdell2004dcq}.

The total lineshape will be a convolution of the inhomogeneous and homogeneous lineshapes. In the simple model in which we take the homogeneous lineshape to be Lorentzian with FWHM $\Delta \nu_\mathrm{L}$ and the inhomogeneous lineshape to be Gaussian with FWHM $\Delta \nu_\mathrm{G}$, the ensemble linewidth can be approximated by $\Delta\nu_\mathrm{L}/2 + \sqrt{\Delta\nu_\mathrm{L}^2/4+\Delta\nu_\mathrm{G}^2}$~\cite{whiting1968eav}. For Al, the narrowest measured homogeneous linewidth is 3.9 GHz with an ensemble linewidth ranging from 4.2-7\,GHz, resulting in an inhomogeneous linewidth ranging from 1.1 to 4.6\,GHz. This range is consistent with the 1.9\,GHz inhomogeneous broadening estimated due to isotope disorder and suggests isotope broadening is the dominant inhomogeneous broadening mechanism.

Both chemical and isotope purification will thus be key to the development of optical quantum technologies with ZnO. Isotope purification, already required to improve the \Do\ spin coherence by removing non-zero spin nuclear isotopes~\cite{linpeng2018cps}, will be further advantageous to reduce broadening from mass disorder. Chemical purity will improve the homogeneous linewidth, either by eliminating impurities/defects that contribute to spectral diffusion and/or lowering the inter-donor spacing to reduce instantaneous spectral diffusion. Even without these materials improvements, the current linewidth properties are sufficient for several applications if donors can be integrated into photonic devices. A Purcell enhancement of less than 100 can enable the generation of indistinguishable photons. Detuned Raman excitation schemes can also mitigate against excited state dephasing~\cite{santori2009odr,he2013its}. Finally, if single donors can be isolated, theoretical schemes to entangle donors with trapped ions~\cite{lilieholm2020pme} should be possible with the current optical properties.

\begin{backmatter}
\bmsection{Funding}
U.S. Department of Energy (DE-SC0020378); National Science Foundation (2212017).

\bmsection{Acknowledgements}
The authors thank Yusuke Kozuka for the bulk ZnO substrates and Simon Watkins for valuable discussions.
This material is based on work primarily supported by the U.S. Department of Energy (DOE), Office of Science, Office of Basic Energy Sciences, under Award No. DE-SC0020378, and partially supported by the National Science Foundation under Grant No. 2212017. 
This work was performed, in part, at the Center for Integrated Nanotechnologies, an Office of Science User Facility operated for the U.S. Department of Energy (DOE) Office of Science. 
Sandia National Laboratories is a multimission laboratory managed and operated by National Technology and Engineering Solutions of Sandia, LLC, a wholly owned subsidiary of Honeywell International, Inc., for the U.S. DOE’s National Nuclear Security Administration under contract DENA-0003525. 
The views expressed in the article do not necessarily represent the views of the U.S. DOE or the United States Government.

\bmsection{Disclosures} 
The authors declare no conflicts of interest.

\bmsection{Data availability}
Data underlying the results presented in this paper are not publicly available at this time but may be obtained from the authors upon reasonable request.

\bmsection{Supplemental document}
See Supplement 1 for supporting content. 

\end{backmatter}

\bibliography{main.bib}

\end{document}

\maketitle

\newpage

\section{Experimental setup and equipment}
\label{app:exp_setup}
\begin{figure}[h]
  \centering
  \includegraphics[width=0.96\linewidth]{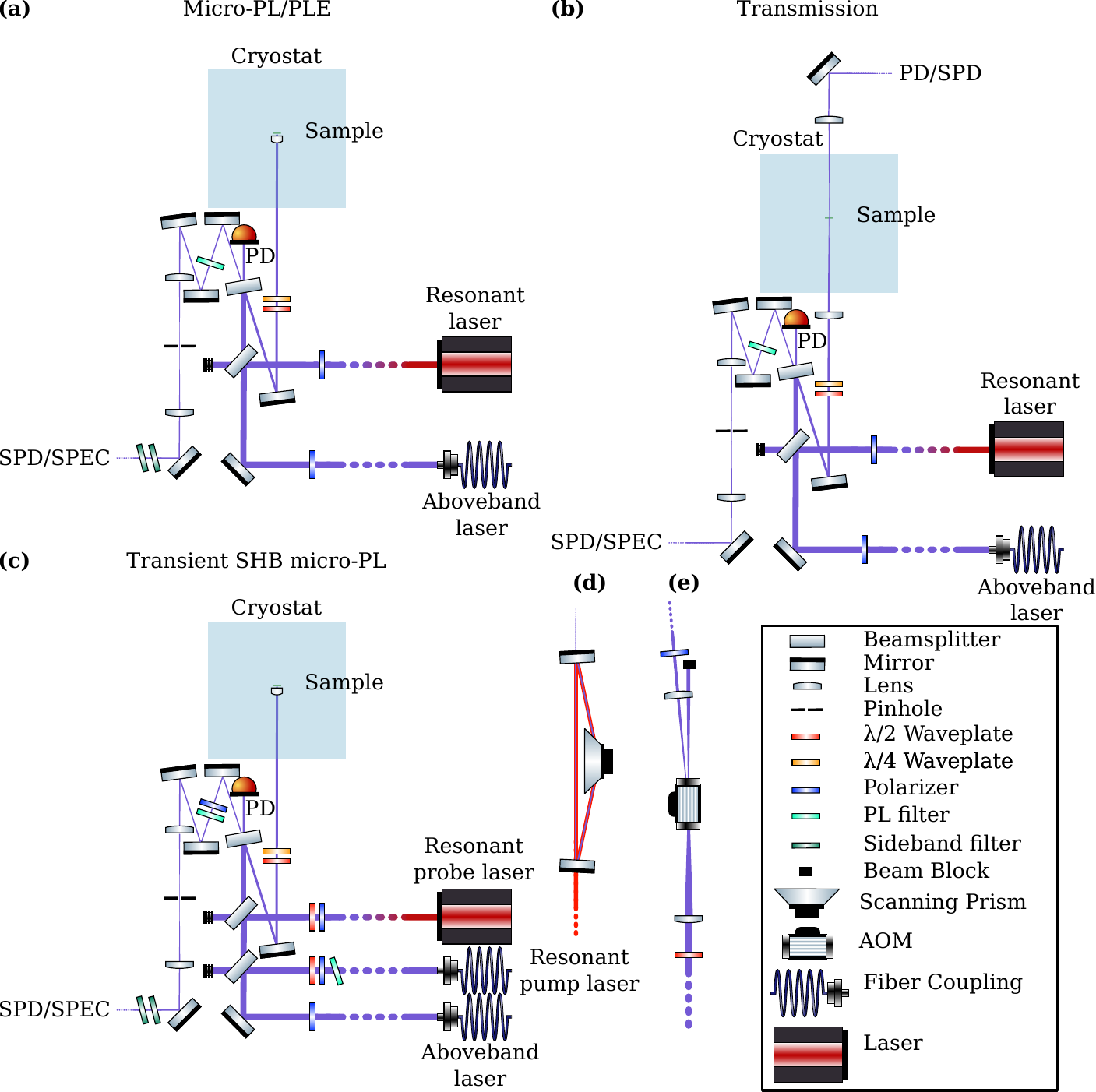}
  \caption{
  \label{fig:opt_path} 
  Here, PD stands for photodiode, SPD for single photon detector, and SPEC for spectrometer. Optical paths for 
  (a) the micro-PL/PLE experiments. This experiment utilizes an aspheric 3.1\,mm-focus lens inside the narrow 1 inch cryostat sample space for PL collection. The emitted (sideband) PL is collected with either an SPD or a spectrometer.
  (b) the transmission experiment. The distinct feature of this experiment is the 200\,mm lens used to excite the sample with a wide beam spot size. The transmitted laser light can be detected by a PD or an SPD, while the emitted PL is collected though the collection path on the front side of the microscope with either an SPD or a spectrometer,
  (c) the transient two-laser micro-PL experiment. This system is very similar to the micro-PL system, but uses two new elements; the second resonant pump laser, and an accousto-optic modulator (AOM) system (see (e)).
  (d) a simplified representation of the unidirectional ring cavity for frequency doubling the resonant probe laser.
  (e) an AOM system to pick-off pulses from the resonant pump and probe lasers.
  }
\end{figure}

Fig.~\ref{fig:opt_path} depicts the optical paths of all three experiments described in the main text. 

In this work, we used three different laser systems for aboveband and resonant excitation. 
For near aboveband excitation we used a CNI MSL-F-360-10mW, a continuous wave (CW) laser emitting near 360\,nm with a maximum power of 10\,mW. 
For resonant excitation and precise frequency scanning control we used a SpectraPhysics Matisse-TS Ti:sapphire laser emitting near 739\,nm.
The Matisse is pumped by a SpectraPhysics Millenia EV CW DPSS green laser emitting at 532\,nm with a maximum power of 15\,W. 
The Matisse is then doubled near 368.5\,nm with a SpectraPhysics WaveTrain frequency doubler, which utilizes a unidirectional ring cavity (see Fig.~\ref{fig:opt_path}d) and yields a conversation efficiency between 4\% and 10\%.
For resonant excitation (pump laser) we use a Toptica DL pro HP laser emitting near 368.9\,nm with a maximum power of 30\,mW. 
The laser emission contains a long tail in the longer wavelength regime, which we cut by using our PL filter.

All of our optics and detectors are graded for use in the near-UV regime. Some notable equipment part numbers are listed below.
% Our mirrors are either MaxMirror (ultrabroadband mirrors) or Thorlabs BB1-E01.
% For 50/50 beam splitters, we either use Thorlabs BSW21, Newport UVBS14-1, or Thorlabs PBS051.
% Achromatic lenses are Newport PAC18AR.15, ashperic lens in cryostate is a Edmund Optics LightPath 354330, and fiber-coupler asheric lenses are Thorlabs C610TME-A, C560TME-A, A397TM-A, 
Our polarizers are either a Thoralabs A-coated Glan-Thompson GTH or a Thoralabs LPUV050.
% The waveplates we use are EKSMA Optics  464-4240 ($\lambda$/2), 464-4440 ($\lambda$/4) and YYY.
The PL filters we use are BrightLine single-band bandpass 370/6\,nm filters, while
the sideband filters we use are BrightLine single-band bandpass 380/14\,nm filters.
Our AOMs are Gooch \& Housego 3307-120 powered by a Gooch \& Housego 1300AF-DIFO-2.5.
% The photodiode we use is XXX.
Our single photon detector is a Laser Components Single Photon Counting Module COUNT BLUE COUNT-50B.
Our spectrometer is an Andor Shamrock 750 with a Netwon DU920P-UVB CCD and a turret with NIR gratings (we use the second order, which is not very efficient, but results in higher resolution).

\section{Excitation power correction}
\label{app:osc_cor}
\begin{figure}[h]
  \centering
  \includegraphics[width=0.5\linewidth]{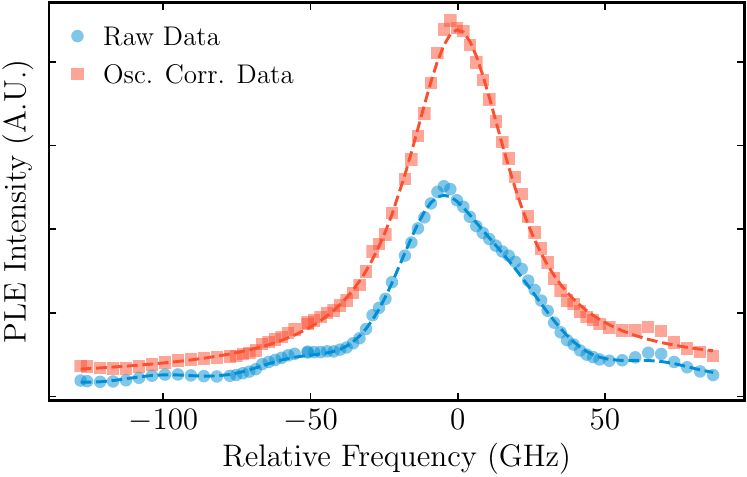}
  \caption{
  \label{fig:ref_corr} 
  PLE of Al at 10.25\,K, sample A. In the raw PLE counts, an oscillation is observed due to an oscillation in the excitation power. The oscillation-corrected-data is obtained by dividing the raw data with $f(E, \phi)$ given in Eq.~\ref{eq:osc_corr_f}.
  }
\end{figure}

An energy dependent modulation of the transmittance, reflectance, and PLE signals is observed. This modulation is an artifact caused by a beamsplitter (BS) in the excitation path of our experimental setup. Reflection from this BS, $P_R$, is incident on the sample. Either transmission through the BS, $P_T$, or the power before the BS, $P_\mathrm{tot}$  is measured to monitor the power in each experiment.   
We find that the ratio $f$ of reflected power divided by the transmitted power fits the following relation,
\begin{equation}
    \label{eq:osc_corr_f}
    f\left(E, \phi \right) = P_R/P_T = c + A\sin\left(2\pi v E + \phi\right)
\end{equation}
with $A=0.07$, $v \simeq \left(0.18\,\,{\rm meV}\right)^{-1}$, $c=0.58$ and a phase $\phi$ which varies between experiments. By normalizing the data by $f$, we are able to remove the oscillations. Example of PLE data before and after the oscillation correction is shown in Fig.~\ref{fig:ref_corr}. 
This method, while effective, may sometimes not completely eliminate the observed oscillations. 
An example of this correction imperfection is the indium optical depth measurement depicted in Fig.~1d of the main text.

\section{Determination of Optical Depth from Transmission}
\label{app:od}
 
We measure total transmission $T$ as a function of wavelength in the frequency window around each \DoX\ resonance. Raw transmission data is corrected for a wavelength-dependent oscillation in the excitation power. This correction is discussed further in Sec.~\ref{app:osc_cor}. Off resonance, the transmission is 0.15 near the Al and Ga peaks and 0.3 near the In peak. Due to this off-resonant absorption, we neglect the effect of reflection from multiple surfaces throughout the measurement window and approximate the total transmission as
\begin{equation}
    T = T_F^2 e^{-\alpha d},
\end{equation}
where $\alpha$ is the absorption coefficient, $d$ is the sample thickness, and the product $\alpha d = \ln(T_F^2/T)$ is the optical depth (OD) plotted in Fig.~1d. $T_F$ is the single face transmission coefficient. To estimate T$_F$, we measure the reflectance between the Al and Ga peaks to be $R = 0.24 \pm 0.02$ and take $T_F \approxeq 1-R$ to be constant over the entire measurement range. 

% \newpage

\section{PL of Sample A in the transmission setup}

\label{app:PLforOD}
\begin{figure}[!h]
  \centering
  \includegraphics[width=0.5\linewidth]{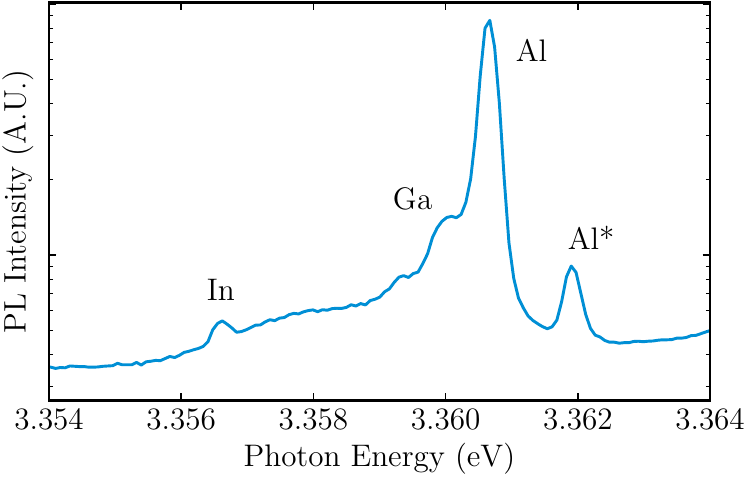}
  \caption{\label{fig:pl_for_od} PL spectrum of sample A. Excitation energy of 3.44\;eV, power of 2.6\;\textmu W, and spot diameter of 380\,\textmu m.  $T = 6.9$\;K, $B=0$\,T.}
\end{figure}

The PL spectrum that corresponds to the transmission experiments in Fig.~1d is shown on Fig.~\ref{fig:pl_for_od}. In addition to being a different sample than that for the spectrum in Fig.~1b, the excitation spot diameter is almost three orders of magnitude larger. In Fig.~\ref{fig:pl_for_od}, the Al emission is $\approx \mathcal{O}\left(10\right)$ stronger than Ga emission which is consistent with the measured OD. 

\section{Donor Density Estimation}
\label{app:density_estimation}

The area under each optical depth (OD) peak in Fig.~1d is proportional to the number of donors in the probed ensemble, and thus can be used to estimate the average donor density.
The donor density of donor $N_x$ ($x$ = Al, Ga, In) can be estimated from~\cite{hilborn1982ecc}
\begin{equation}
\label{eq:donor_density}
    N_x= 8\pi \frac{g_{\Do}}{g_{\DoX}} \left(\frac{n}{\lambda_x}\right)^2\tau_{x, \mathrm{rad}}\int \alpha(\nu) d\nu ,
\end{equation}
where $g_i$ is the degeneracy of state $i$, $n$ is the index of refraction, $\lambda_x$ is the vacuum transition wavelength, $\tau_{x, \mathrm{rad}}$ is the zero-phonon-line radiative lifetime. 
We estimate the ZPL radiative lifetime as the total radiative lifetime determined by the slow decay component of the experimental lifetime, divided by the fraction of emission into the ZPL. This fraction is determined by the Huang-Rhys parameter~\cite{wagner2011bez} and the ratio of the two-electron satellite transitions to the phonon-assisted transitions. 
We obtain $\tau_{x,rad} = 0.95$\,ns, 1.18\,ns, and 1.52\,ns for $x=$ Al, Ga, and In, respectively. Using Eq.~\ref{eq:donor_density}, we find $N_{\rm Al} = 7.5\cdot 10^{15} \rm\, cm^{-3}$, $N_{\rm Ga} = 9.9\cdot 10^{14} \rm\, cm^{-3}$, and $N_{\rm In} = 7.4\cdot 10^{13} \rm\, cm^{-3}$. These values, measured for Sample A, are within an order of magnitude of the SIMS values measured for sample B. The relative concentrations are consistent with the PL spectra for this sample (Sec.~\ref{app:PLforOD}).

\section{Temperature dependence of implanted In PLE linewidth}
\label{app:implanted_in_temp_dep}

\begin{figure}[h]
  \centering
  \includegraphics[width=0.5\linewidth]{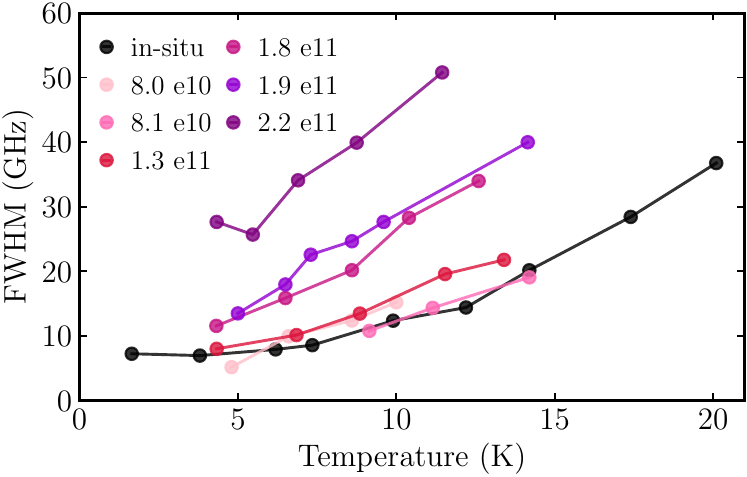}
  \caption{\label{fig:temp_dep_implanted} Temperature-dependent linewidth of the In \DoX\ transition in Sample B for different implantation fluences. The legend provides the fluences in units of cm\textsuperscript{-2}.}
\end{figure}

Sample B provided an opportunity to measure the PLE linewidth as a function of temperature for different In implantation fluences (Fig.~\ref{fig:temp_dep_implanted}). At the lowest studied fluences ($< 1.3\cdot 10^{11} $\,cm\textsuperscript{-2}) the dependence of linewidth on temperature closely matches the \textit{in-situ}-doped In in sample A. This lowest implantation density is estimated to be $\sim$50 times higher than the {\it in-situ}-doped density estimated in Sec.~3.1. This similarity supports that the temperature dependence described in Sec.~3.2 is intrinsic to the donor, and independent of environmental damage or defect density. As implantation dose increases further, however, both a broader linewidth and a steeper dependence of the linewidth on temperature is observed. The mechanism for the steeper dependence is unknown, but could be due to relaxation from the \DoX\ state to defect-states created during the implantation process.

\section{\texorpdfstring{\DoXs}~~Magneto-PL for Al, Ga, and In}
\label{app:DoXs_magneto_PL}

\begin{figure}[h]
  \centering
  \includegraphics[width=0.5\linewidth]{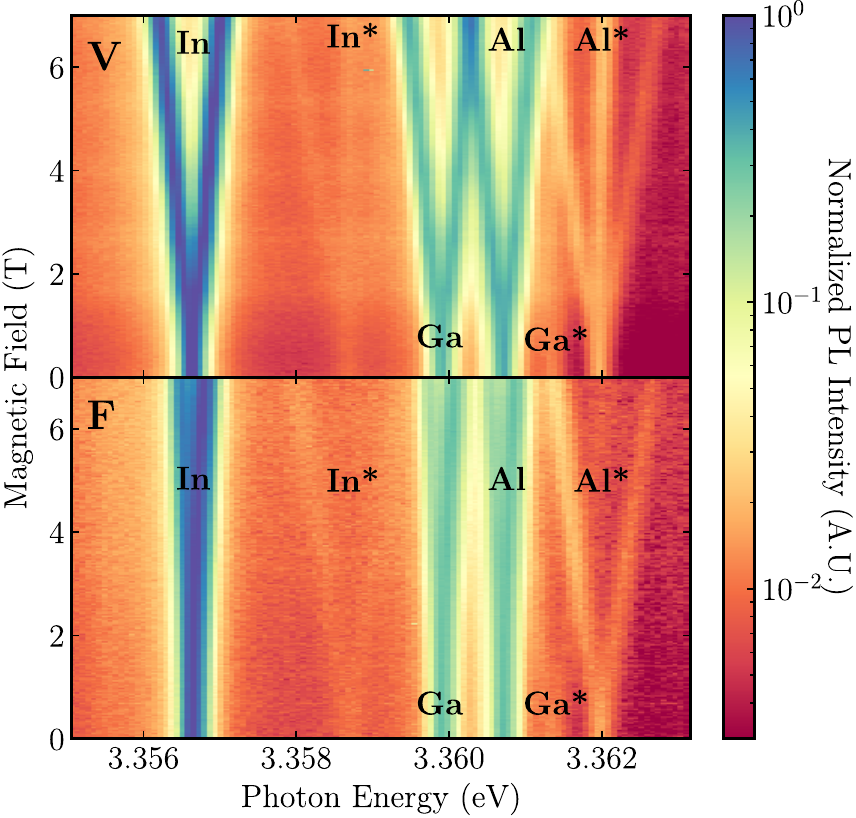}
  \caption{\label{fig:DoXs_magneto_PL} Magnetic field dependence of \DoX\ transitions for all donors in sample B, at 7.2\;K, with excitation energy 3.44\;eV and power 200\;nW.}
\end{figure}

The magneto-PL spectra for all three donors is shown in Fig.~\ref{app:DoXs_magneto_PL}. Unlike sample A, sample B has a substantial density of In, which facilitates the study of the weak In \DoXs\ line.
The Al \DoXs\ magnetic field dependence is identical in Sample A and B. The In \DoXs\ has a similar magnetic field dependence to Al, splitting into three lines in Voigt and into two lines in Faraday geometry. 
Unfortunately, the Al \DoX\ emission obscures the Ga \DoXs\ dependence on magnetic field. However, we can observe that the Ga \DoXs\ splits in two lines in the Faraday geometry, while the transition that is independent of the magnetic field is clearly visible in the Voigt geometry.

\section{\texorpdfstring{\Do}~~and \texorpdfstring{\DoX}\ ~State Models}
\label{app:state_models}

The models of the \Do\ and \DoX\ state effective mass envelope functions presented here are used to calculate the effect of the local nuclear spin and isotopic environments on the \transX\ transition. Both the neutral donor and donor-bound exciton effective mass envelope functions are assumed to be isotropic, which is a reasonable approximation as the dielectric constant, electron effective mass, and hole effective mass are all close to isotropic in ZnO~\cite{meyer2004bed}. The same isotropic assumption is made for Si. Si carrier effective masses are determined by conduction measurements~\cite{riffe2002tds}. The effective mass theory parameters used to calculate the carrier envelope functions are listed in Tables~\ref{table:ZnO} and \ref{table:Si}, where $m_0$ is the free electron mass, and $\epsilon_0$ is the permittivity of free space.

\begin{table}[h]
\centering
\begin{tabular}{cc}
 
 \hline
 Constant & Value \\
 % \hline 
 \hline
 electron effective mass & $m_{e}$ = 0.27 $m_0$~\cite{meyer2004bed, rossler1999iic} \\
 % \hline 
 hole effective mass & $m_{h}$ = 0.59 $m_0$ \cite{meyer2004bed, rossler1999iic}  \\
 % \hline 
 dielectric constant & $\epsilon_{\rm ZnO}$ = 8.2 $\epsilon_0$ \cite{meyer2004bed, rossler1999iic}  \\
 % \hline 
 Al \Do\ binding energy & $E_{\rm Al}$ = 51.5 meV \cite{wagner2011bez} \\
 % \hline 
 Ga \Do\ binding energy & $E_{\rm Ga}$ = 54.6 meV \cite{wagner2011bez}  \\
 % \hline 
 In \Do\ binding energy & $E_{\rm In}$ = 63.2 meV \cite{wagner2011bez}  \\
 
 \hline
 
\end{tabular}
\caption{ZnO effective mass theory parameters}
\label{table:ZnO}
\end{table}

\begin{table}[h]
\centering
\begin{tabular}{cc}

 \hline
 Constant & Value \\
 \hline 
 % \hline
 electron effective mass & $m_{e}$ = 0.26 $m_0$ \cite{riffe2002tds} \\
 % \hline 
 hole effective mass & $m_{h}$ = 0.33 $m_0$ \cite{riffe2002tds}  \\
 % \hline 
 dielectric constant & $\epsilon_{\rm Si}$ = 11.7 $\epsilon_0$ \cite{dunlap1953dmd}  \\
 % \hline 
 P \Do\ binding energy & $E_{\rm P}$ = 45.59 meV \cite{jagannath1981lee} \\
 
 \hline
 
\end{tabular}
\caption{Si effective mass theory parameters}
\label{table:Si}
\end{table}

\subsection*{\texorpdfstring{\Do}~~Model}

The neutral donor model presented here follows Ref.~\cite{heine1975tis}. The \Do\ electron radial envelope function is given by
\begin{equation}
    \Psi_{D^0,e} = \left(\frac{2^{2n}}{2 n a^3 \Gamma(2n + 1)} \right)^{1/2} r^{n-1} e^{-r/a}
    \label{eq:dowavefunction}
\end{equation}
where
\begin{equation}
    a = \left( \frac{\hbar^2}{2 m_e E_b} \right)^{1/2}, \quad \text{and }\quad n = \left( \frac{E_H}{E_b} \right) ^{1/2}.
\end{equation}
Here, $\Gamma$ is the gamma function, $m_e$ is the electron effective mass, $E_b$ is the donor binding energy, and $E_H$ is the hydrogenic effective mass binding energy. The hydrogenic binding energy is given by 
\begin{equation}
    E_H = \frac{m_e}{m_0} \frac{\epsilon_0^2}{\epsilon_{ZnO}^2} R_y,
\end{equation}
in which $m_0$ is the free electron mass, $\epsilon_0$ is the vacuum permittivity, and $R_y$ is the Rydberg energy. Eq.~\ref{eq:dowavefunction} assumes the hydrogenic approximation is nearly correct, but includes distinct central cell corrections between different impurity elements. 

\subsection*{\texorpdfstring{\DoX}~~Model}

A simple model for the bound exciton is proposed in Ref.~\cite{puls1983esb}, in which the two electrons in the bound exciton reside close to the positively-charged impurity while the positive hole binds to the net negative system consisting of the two electrons and the impurity. The potential binding the hole is given by
\begin{equation}
    V(r_h) = - \frac{e_c^2}{4\pi\epsilon_{ZnO} r_h}[1 - 2e^{-2r_h / a_e} (1 + r_h/a_e)].
    \label{eq:truePotent}
\end{equation}
Here $r_h$ is the radial position of the hole, while $a_e$ is the effective Bohr radius for the two electrons. The two electrons in the bound exciton are assumed to occupy the same 1-s orbital state. The resulting hole potential is not easily solvable, but can be approximated by the Kratzer potential,
\begin{equation}
    V(r_h) \approxeq - 2D \left( \frac{b}{r_h} - \frac{b^2}{2r_h^2}\right),
    \label{eq:kratzer}
\end{equation}
with $D = s e_c^2 / 8\pi\epsilon_{ZnO} a_e$, and $b = t a_e$, where $t$ and $s$ are fitting constants for the best fit of the Kratzer potential Eq.~\ref{eq:kratzer} to Eq.~\ref{eq:truePotent}, with $t = 1.337$, $s = 1.0136$~\cite{puls1983esb}. The solution to the Kratzer potential is calculated in Ref.~\cite{bayrak2007eas}. For modeling the ground state we set the rotational and radial quantum numbers $l_h$ = $n_h$ = 0. With these assumptions, the hole's radial envelope function in the bound exciton ground state is
\begin{equation}
    R(r) = r^{\Lambda_{00}} \text{exp} (-\epsilon r),
\end{equation}
in which, $\epsilon = \sqrt{-2m_h E_{h,00}/\hbar^2}$. The hole energy, $E_{h, n_h l_h}$, and factor $\Lambda_{n_h l_h}$ are defined as follows.
\begin{equation}
    \label{eq:hole_energy}
    E_{h, n_h l_h} = -\frac{(2 m_h b^2/\hbar^2) D^2}{ \left(1 + \Lambda_{00} \right)^{-2}}, \Lambda_{n_hl_h} = -\frac{1}{2} + n_h + \sqrt{\left(l_h + \frac{1}{2}\right)^2 + \frac{2 m_h b^2 D}{\hbar^2}}
\end{equation}
The total energy of the bound exciton, obtained from \cite{puls1983esb}, is Eq.~\ref{eq:aeEnergyPuls}, where the effective Bohr radius of the bound electrons is allowed to vary. The final term of Eq.~\ref{eq:aeEnergyPuls} is the hole energy, Eq.~\ref{eq:hole_energy}. To obtain the bound exciton ground state, Eq.~\ref{eq:aeEnergyPuls} is minimized with respect to the bound electron radius $a_e$, with hole quantum numbers $n_h$ = $l_h$ = 0. 
\begin{equation}
\begin{split}
    E(a_e) = 2E_g + 2R_D\left[\left(\frac{a_D}{a_e}\right)^2 - \frac{11}{8}\frac{a_D}{a_E}\right] \\ - 
    2R_D\left[\frac{s^2 t^2}{2} \frac{m_h}{m_e} \left(n_h + \frac{1}{2} + \sqrt{\left(l_h + \frac{1}{2} \right) + \frac{st^2 a_e}{a_D}\frac{m_h}{m_e}}  \right)^{-2} \right]
    \label{eq:aeEnergyPuls}
\end{split}
\end{equation}

Here $a_D$ is the effective Bohr radius of the neutral donor. To account for the central cell differences between donors in the \Do\ model, this is determined for each donor by $a_D = 2\langl r \rangl/3$, which is exact for perfectly hydrogenic 1s states. The resulting effective Bohr radii of the D$^0$X electrons are $a_{e\textrm{,Al}}$ = 2.08 nm, $a_{e\textrm{,Ga}}$ = 1.98 nm, $a_{e\textrm{,In}}$ = 1.75 nm, which result in values for $b$ in Eq.~\ref{eq:kratzer} of $b_{\textrm{Al}}$ = 2.8 nm, $b_{\textrm{Ga}}$ = 2.6 nm, $b_{\textrm{In}}$ = 2.3 nm. For the phosphorus donor in silicon, we obtain $a_{e\textrm{,P}}$ = 1.95 nm and $b_{\textrm{P}}$ = 2.6 nm.

\section{Discussion of \texorpdfstring{\DoXs}~~States}
\label{app:DoXs_discussion}

A model is proposed by Meyer {\it et al.} in Ref.~\cite{meyer2004bed, meyer2010esp} to explain the excited states of the bound exciton, including those discussed in Sec.~3.2. They introduce two components to this model, one for the lower energy excited states near 2 meV from the ground state, and another for the series of excited states found above 6 meV. Ref.~\cite{meyer2010esp} explains the excited states found around and above 6 meV by electronic excited states of the hole, given by increasing $n_h$ and $l_h$ in Eq.~\ref{eq:aeEnergyPuls} and minimizing with respect to $a_e$. The values for these excited states obtained from our model in Sec.~\ref{app:state_models} agree well with experimental values presented in Table III of Ref.~\cite{meyer2010esp}.

Ref.~\cite{meyer2010esp} attributes the set of energies found at around 2 meV to rotational-vibrational states of the bound exciton. Ref.~\cite{meyer2010esp} evaluates these energies using the Kratzer potential, as given in Eq.~\ref{eq:kratzer}, arriving at the following equation for the rotational-vibrational energies $E(\nu, J)$.

\begin{equation}
    \label{eq:ro-vib}
    E(\nu, J) = -\frac{(2mb^2/\hbar^2)D^2}{\left[\left(\nu + \frac{1}{2}\right) + \sqrt{\left(J+\frac{1}{2}\right)^2 + \left(\frac{2mb^2}{\hbar^2}\right)D}\right]^2}
\end{equation}
In analogy to rotational-vibrational states of diatomic molecules ({\it i.e.} as in Ref.~\cite{pliva1999cre}), Ref.~\cite{meyer2010esp} interprets $D$ as the bound exciton localization energy, $m$ as the effective mass of the hole, and $a$ as the distance between the hole and impurity. While Meyer \textit{et al.}'s interpretation of $a$ and $D$ yield excited states matching those observed around 2\,meV, to determine $b$, Ref.~\cite{meyer2010esp} introduces a \textit{pseudo-acceptor} model approximation for the donor bound exciton, in which the electrons are tightly bound to the impurity and the hole orbits the resulting negatively charged center at a distance equal to the approximate effective Bohr radius of the acceptor in ZnO, 0.8 nm. The bound exciton model in App~\ref{app:state_models} assumes the electrons are nearer to the impurity than the hole similarly to our approach in Sec.~\ref{app:state_models}. However, the energy of the system Eq.~\ref{eq:aeEnergyPuls} is minimized when $b \approxeq$ 2.3-2.8\,nm, far from 0.8\,nm, implying that the \textit{pseudo-acceptor} approximation is inappropriate. Meyer \textit{et al.}'s interpretation also deviates from the original source of this equation \cite{ruhle1978ebn}, where $D$ is the minimum of the Kratzer potential binding the hole and $a$ is the radial position of this minimum. Following the source interpretation of Eq.~\ref{eq:ro-vib}, Eq.~\ref{eq:ro-vib} is equal to the final term of Eq.~\ref{eq:aeEnergyPuls}, and both predict excited states above 6\,meV. Eq.~\ref{eq:ro-vib} and Eq.~\ref{eq:aeEnergyPuls} are only distinguished by the accuracy of their prediction of these states, with Eq.~\ref{eq:aeEnergyPuls}'s improved accuracy due to its inclusion of changes in $a_e$ between excited states of the hole. The interpretation by Meyer \textit{et al.} in analogy to rotational-vibrational states of diatomic molecules also assumes the g-factor of these excited states is equal to that of the main line, which is contradicted by our results in Sec.~3.2 and Sec.~\ref{app:DoXs_magneto_PL}.  

\section{Effect of the Nuclear-Spin Environment }

The nuclear-spin environment primarily affects the \Do\ levels in the \transX\ transition via the electron-nuclear contact hyperfine interaction.
The strength of the contact hyperfine interaction for \DoX\ will be much smaller as the two electrons form a spin-0 singlet state and the hole is predominantly p-type~\cite{preston2008bsz}. The \Do -bound electron interacts with both the lattice nuclear-spins and the donor's nucleus.

The broadening due to the non-zero \ZnwNS\ lattice spins is given by a Gaussian distribution of the hyperfine field, $\exp\left({-B^2/\Delta^2_{B,Zn}}\right)$, and has a dispersion~\cite{merkulov2002esr,linpeng2018cps}
$$
\Delta_{B,Zn} = \frac{\mu_0\mu_{\mathrm{Zn}}}{g_e}\sqrt{\frac{32}{27}}\sqrt{\frac{I_\mathrm{Zn}+1}{I_\mathrm{Zn}}}|u_\mathrm{Zn}
|^2 \sqrt{f\sum_i|\Psi(\vec{R}_i)|^4},$$
in which $g_e$ is the electron g-factor, $\mu_0$ is the vacuum permeability, $I_\mathrm{Zn} = 5/2$ is the \ZnwNS\ nuclear spin, $\mu_\mathrm{Zn} = 0.874\,\mu_N$ is the \ZnwNS\ nuclear magnetic moment in terms of the nuclear magneton $\mu_N$, $f=4.1\%$ is the natural abundance of \ZnwNS, $\Psi{(\vec R_i)}$ is the effective mass wavefunction at the $i$th Zn lattice site, and $|u_\mathrm{Zn}|^2$ is the ratio of the Bloch function density at the Zn site to the average Bloch function density. Ref.~\cite{linpeng2018cps} uses a pure hydrogenic effective mass wavefunction for $\Psi(R)$ with Bohr radius of 1.7\,\AA\ to estimate a 22\,MHz linewidth. Using a slightly modified $\Psi(R)$\  which accounts for the central cell correction due to the three different types of donor (Sec.~\ref{app:state_models}), we obtain \ZnwNS\ nuclear-spin broadened lines of 22, 24 and 29\,MHz for Al, Ga and In, respectively. These values are only slightly larger than the reported optically-detected magnetic resonance linewidths for Ga and In of 19 and 22 MHz respectively~\cite{gonzalez1982mrs}. 

The hyperfine splitting due to the donor nuclear spin of spin $I$ depends on the ratio of the Zeeman splitting $g\mu_B B$ to the hyperfine constant $A$~\cite{mohammady2012aqc}. 
At 0-field, the $2(2I+1)$ states split into two sublevels separated by $A\sqrt{\frac14 +I(I+1)}$. 
At high magnetic field, $g_e\mu_BB \gg A$, each electron Zeeman level splits into (2I + 1) lines with a splitting of $A/2$ between hyperfine levels. 
$A_{\mathrm{Al}}=1.45$\,MHz~\cite{orlinskii208isa}, $A_{\mathrm{Ga}}=11.5$\,MHz~\cite{gonzalez1982mrs}, and $A_{\mathrm{In}}= 100$\,MHz~\cite{gonzalez1982mrs, block1982odm} with $I_{\mathrm{Al}} = 5/2$, $I_{\mathrm{Ga}} = 3/2$ and $I_{\mathrm{In}}=9/2$. 

The nuclear-spin environment contributes to the inhomogeneous broadening of the \transX\ transition linewidth.
However, any experiment conducted over timescales longer than the relaxation time of the nuclear-spin states will be affected by this broadening. Comparing to the several GHz linewidths observed in Fig.~4d, the nuclear-spin contribution is negligible. However, it is the main source of donor ground-state dephasing~\cite{linpeng2018cps}, and significantly affects the two-laser linewidth in coherent population trapping experiments~\cite{wang2023pdq, viitaniemi2022csp}. For In, the hyperfine splitting is comparable to the lifetime-limited linewidth and could theoretically be optically resolved. 

\section{Impurity Isotope Effect}

The shift $\Delta E^{don}_{S,c}$ for state S and carrier c due to substitution of the donor atom isotope is given by Heine and Henry in Ref.~\cite{heine1975tis} as 
\begin{equation}
    \Delta E^{don}_{S,c} =
    \frac{2 \hbar \omega_D}{5}\left(\frac{M_0}{M}\right)^{1/2} \frac{\Delta M}{M} \frac{\gamma_c}{\gamma} \left(-\frac{dE_g}{dkT}\right)_{HT} P_{S,c}
    \label{eq:donorisotope}
\end{equation}
in which $\omega_D$ is the Debye frequency of ZnO, 
$M_0$ is the average substituted atom (Zn) mass, 
$M$ is the mass of the lightest donor isotope, 
$\Delta M$ is the difference between the mass of the heavier and lightest donor isotope, and
$\left(\sfrac{dE_g}{dkT}\right)_{HT}$ is the temperature dependence of the band gap at high temperature.
These values are provided in Table~\ref{table:ZnOdonorAtom} in Sec.~\ref{app:isotope_shift}. 
$\gamma_c$ is the fractional oscillator force constant reduction due the presence of a carrier $c$,
$\gamma$ = $\gamma_e$ + $\gamma_h$, is the sum of the force reduction for holes and electrons, with $\gamma_h$ = 3$\gamma_e$~\cite{heine1975tis}, and $P_{S, c}$ is the average volume per atom multiplied by the average amplitude squared within a sphere encompassing the donor and its nearest neighbors (0.2 nm in ZnO).

\begin{table}[h]
\centering
\begin{tabular}{cc}
 \hline 
 Constant & Value \\
 \hline 
 % \hline
 $\hbar\omega_d$ (meV) & 35.8 ($T_D$ = $416$\,K) \cite{rossler1999iic} \\
 % \hline 
 Ga: $M_0$, $M$, $\Delta M$ (amu) & 69, 71, 2 \cite{cardona2005ieo}  \\
 % \hline 
 In: $M_0$, $M$, $\Delta M$ (amu) & 113, 115, 2 \cite{cardona2005ieo} \\
 % \hline 
 $\left(\sfrac{dE_g}{dkT}\right)_\text{HT}$ (meV/K) & 3.24 at HT 400\,K \cite{rossler1999iic} \\
 
 \hline
 
\end{tabular}

\caption{ZnO donor atom isotopic substitution constants}
\label{table:ZnOdonorAtom}
\end{table}

The shift from impurity atom isotopic substitution is determined by Eq.~\ref{eq:donorisotope} with constants given in Table~\ref{table:ZnOdonorAtom}. Both aluminum and phosphorus have only one commonly occurring isotope and are thus not included in Table~\ref{table:ZnOdonorAtom}. Impurity isotope substitution shifts the \transX\ transition energy by only 16 MHz between Ga$^{69}$ and Ga$^{71}$ and 13 MHz between In$^{113}$ and In$^{115}$, implying that variation in Zn and O isotopes in the defect environment are the dominant isotopic broadening effect. 

\section{Isotopic Environment Simulation}
\label{app:isotope_shift}

\begin{table}[h]
\centering
\begin{tabular}{ccc}

 \hline
 Energy, $W_{i,c}$ & Atom/Isotope & Rel. Abundance \\
 \hline 
 \hline
 \multicolumn{3}{c}{Valence Band, $W_{i,h}$} \\
 \hline
 \hline
 0 meV & $^{64}$Zn & 48.6 $\%$ \\
 % \hline
 0.66 meV & $^{66}$Zn & 27.9 $\%$\\
 % \hline
 0.98 meV & $^{67}$Zn & 4.1 $\%$\\
 % \hline
 1.31 meV & $^{68}$Zn & 18.8 $\%$\\
 % \hline
 1.97 meV & $^{68}$Zn & 0.6 $\%$ \\
 % \hline
 \hline
 0 meV & $^{16}$O & 99.75 $\%$ \\
 % \hline
 2.50 meV & $^{17}$O & 0.05 $\%$\\
 % \hline
 5.00 meV & $^{18}$O & 0.2 $\%$\\
 % \hline
 
 \hline 
 \hline
 \multicolumn{3}{c}{Conduction Band, $W_{i,e}$} \\
 \hline
 \hline
 0 meV & $^{64}$Zn & 48.6 $\%$ \\
 % \hline
 0.16 meV & $^{66}$Zn & 27.9 $\%$\\
 % \hline
 0.25 meV & $^{67}$Zn & 4.1 $\%$\\
 % \hline
 0.33 meV & $^{68}$Zn & 18.8 $\%$\\
 % \hline
 0.49 meV & $^{68}$Zn & 0.6 $\%$ \\
 % \hline
 \hline
 0 meV & $^{16}$O & 99.75 $\%$ \\
 % \hline
 0.65 meV & $^{17}$O & 0.05 $\%$\\
 % \hline
 1.25 meV & $^{18}$O & 0.2 $\%$\\
 \hline
 
\end{tabular}

\caption{Isotopic perturbation energies in ZnO. Isotopic abundances are obtained from~\cite{cardona2005ieo}.}
\label{table:ZnOPerturb}
\end{table}

For Al, Ga, and In defects in ZnO, the isotopic environment is simulated up to 10\,nm from the defect, which is sufficient to include the \Do\ and \DoX\ states. The shifts $W_{i,c}$ from environmental zinc and oxygen are calculated using the dependence of the donor lines on isotopic substitution, which is measured in Ref.~\cite{manjon2003zws} to be $dE_{\DoX}$/$dM_{Zn} = 0.41 \pm 0.05$\,meV/amu for zinc, and $dE_{\DoX}$/$dM_{O} = 3.12 \pm 0.05$\,meV/amu for oxygen. Adopting our 80$\%$ valence and 20$\%$ conduction band shift assumption~\cite{gorkavenko2007ctd}, we obtain 
\begin{equation}
    W_{i,c} = S_{c} \Delta M \left(\frac{dE_{D^0X}}{dM}\right).
    \label{eq:isoShiftDep}
\end{equation}
Here $S_c$ is the fraction of the total band gap shift due to the valence band or conduction band; if the carrier in question is a hole, the energy shift will be due to the shift in the valence band, if it is an electron, the shift will be due to the shift in the conduction band ($S_h$ = 0.8, $S_e$ = 0.2). $\Delta M$ is the difference between the lightest isotope and a given environmental isotope in amu, and $dE_{\DoX}$/$dM_{elem}$ is the environmental mass dependence of the $D^0X$ transition energy for element $elem$. The values for environmental isotopic perturbation in ZnO as obtained from Eq.~\ref{eq:isoShiftDep} are given in Table~\ref{table:ZnOPerturb}. 

\begin{table}[!ht]
\centering
\begin{tabular}{ccc}
 
 \hline
 Energy, $W_{a,b}$ & Atom/Isotope & Rel. Abundance \\
 
 \hline 
 \hline
 \multicolumn{3}{c}{Valence Band, $W_{i,h}$} \\
 \hline
 \hline
 0 meV & $^{28}$Si & 92.2 $\%$ \\
 % \hline
 0.76 meV & $^{29}$Si & 4.7 $\%$\\
 % \hline
 1.52 meV & $^{30}$Si & 3.1 $\%$\\
 % \hline
 
 \hline 
 \hline
 \multicolumn{3}{c}{Conduction Band, $W_{i,e}$} \\
 \hline
 \hline
 0 meV & $^{28}$Si & 92.2 $\%$ \\
 % \hline
 0.26 meV & $^{29}$Si & 4.7 $\%$\\
 % \hline
 0.52 meV & $^{30}$Si & 3.1 $\%$\\
 \hline
 
\end{tabular}

 \caption{Isotopic perturbation energies in Si. Isotopic bundances are obtained from~\cite{cardona2005ieo}.}
 \label{table:SiPerturb}
\end{table}

For the phosphorus defect in Si, the isotopic environment is simulated up to 12 nm (as it is slightly shallower). The breakdown of the band gap shift between the conduction and valence band shifts is noted in~\cite{karaiskaj2003ora} as 75$\%$ valence band and 25$\%$ conduction band. $dE_{\DoX}$/$dM_{Si}$ = 1.02\,meV, which is obtained from~\cite{cardona2005ieo}. From Eq.~\ref{eq:isoShiftDep}, we find the isotopic perturbation energies shown in Table~\ref{table:SiPerturb}.

\begin{figure}[h]
  \centering
  \includegraphics[width=0.5\linewidth]{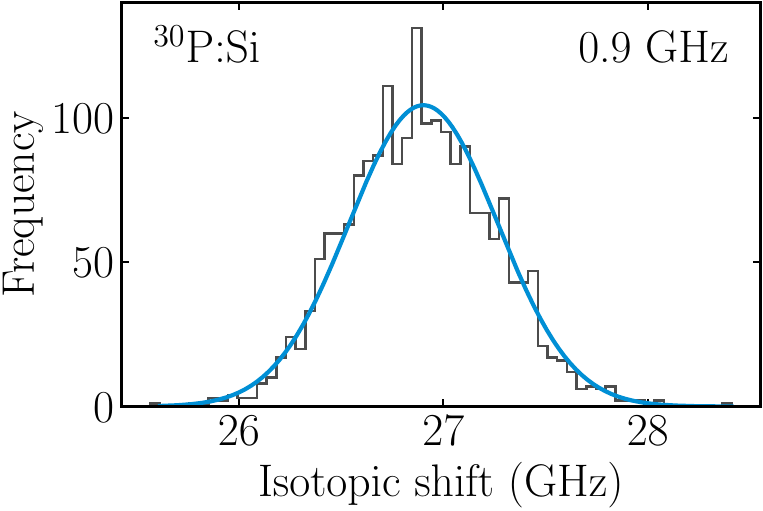}
  \caption{\label{fig:silicon_isotope} Inhomogeneous broadening simulation for Silicon at 2000 simulated environments. The resulting inhomogeneous broadening is 0.9 GHz. This compares well to the measured isotopic inhomogeneous broadening of 1.1 GHz \cite{yang2009hlp}.}
\end{figure}

The isotopic environment's contribution to the linewidth of the phosphorus shallow donors in Si's bound exciton transition has been measured in high quality natural silicon~\cite{yang2009hlp} as 1.1\,GHz.
To verify our model we simulated this transition, yielding a broadening of 0.9\,GHz (see Fig.~\ref{fig:silicon_isotope}), in good agreement with the experimentally observed value.
The primary source of discrepancy is likely found in the simplified \DoX\ state model. Given the difficulty in obtaining an accurate model of the donor bound exciton state, such discrepancies are expected. Despite the uncertainty regarding the \DoX\ state model, the agreement between the model and the experimental data supports the validity of our approach. 

\section{Investigating power broadening of spectral anti-hole burning}
\label{app:rshb_power_dep}
\begin{figure}[h]
  \centering
  \includegraphics[width=0.5\linewidth]{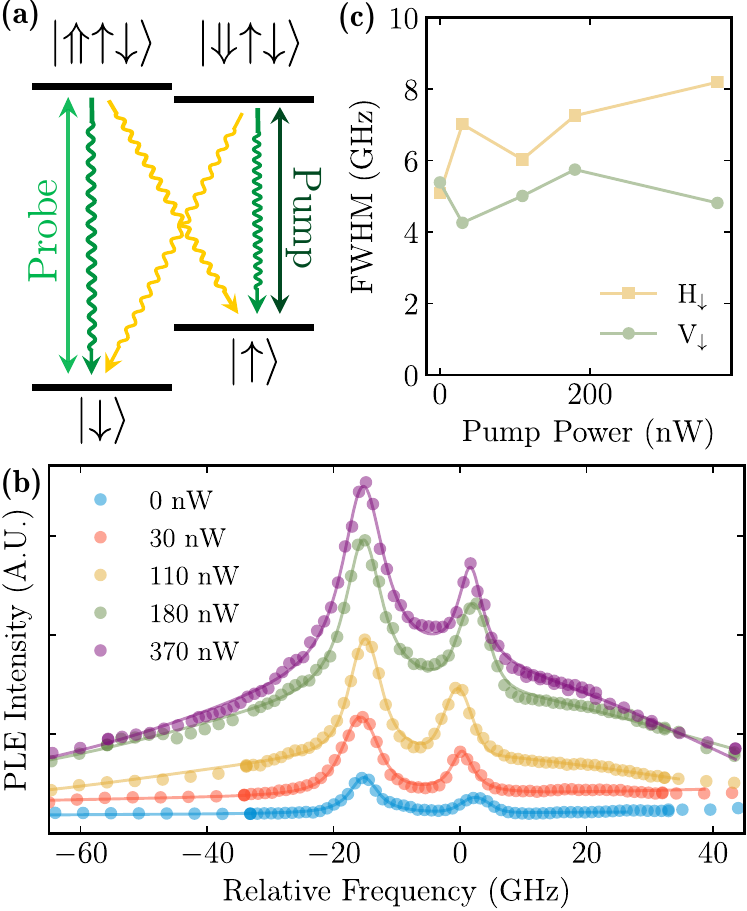}
  \caption{\label{fig:rSHB_voigt} 
  (a) Energy diagram for a pump-probe experiment in Voigt geometry.
  (b) PLE spectra of Al \DoX\ at 1.8\,K and 7\,T of different pump powers. The probe power is held constant at 290\,nW. 
  (c) The anti-hole linewidth for the $V_\downarrow$ and $H_\downarrow$ transitions as a function of pump power.
  }
\end{figure}
We additionally performed continuous-wave spectral anti-hole burning experiments as a function of pump pulse power. These measurements are performed in the Voigt geometry at 6\,T and 1.8\,K. The pump laser is resonant on the $V_\uparrow$ transition, while the probe laser is scanned near the $V_\downarrow$ and $H_\downarrow$ transitions as shown in Fig.~\ref{fig:rSHB_voigt}a. In this geometry, two anti-holes are observed, split by the hole Zeeman factor (Fig~\ref{fig:rSHB_voigt}b). As shown in Fig~\ref{fig:rSHB_voigt}c, the anti-hole linewidth does not depend on pump power. These results indicate the anti-hole linewidth is already maximally broadened by the lowest probe intensities in these steady-state experiments and/or is dominated by the 290\,nW probe laser. 

\section{Delay dependence of two-laser transient spectroscopy}
\label{app:delay_dep}

\begin{figure}[h]
  \centering
  \includegraphics[width=0.5\linewidth]{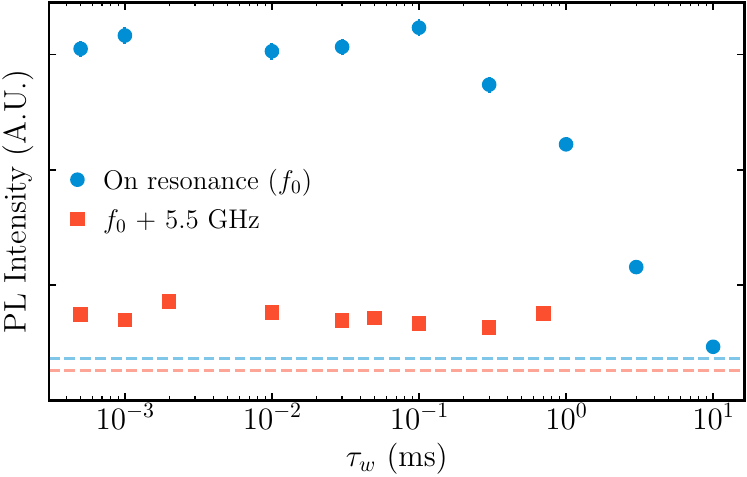}
  \caption{\label{fig:delay_dep} 
  PL intensity at the start of the probe pulse during pump-probe experiment, as a function of wait time between lasers $\tau_w$ at 1.7\;K and 7\;T. The 170\,nW pump laser was on for 100\;\textmu s, and the 230\,nW probe laser was on for 1\;ms. The semi-transparent dashed lines correspond to the optically-pumped intensity at the end of the 1 ms probe pulse. $f_0$ denotes the center frequency of the Al \DoX\ emission.
  }
\end{figure}

We performed additional transient pump-probe experiments as a function of wait time $\tau_w$ between the pump and probe lasers, for two different probe excitation energies, on-resonance with the Al \DoX\ peak, and +5.5\,GHz off-resonance.
Fig.~\ref{fig:delay_dep} depicts the PL intensity integrated over the first two microseconds of the probe pulse. 
We observe the same PL intensity for wait times smaller than 0.1\;ms, while,
After 0.1\;ms, the population of the \down\ state starts to deplete due to the longitudinal spin relaxation ($T_1 = 1.5$\;ms), as described in~\cite{niaouris2022esr}.
The near-constant PL intensity over wait times shorter than \Tone\ indicates that the process governing the homogeneous broadening does not occur while the probe laser is off, but occur under optical excitation. 

\newpage
\bibliography{supplement.bib}